\documentclass[amsmath,amssymb,twocolumn,aps,prl,footinbib,longbibliography]{revtex4-1}

\usepackage[colorlinks=true,linkcolor=blue,urlcolor=blue,citecolor=blue]{hyperref}

\newcommand{\red}{}

\newcommand{\blue}{}

\usepackage{times}

\usepackage[normalem]{ulem}
\usepackage{tikz}
\usetikzlibrary{arrows,shapes,positioning,snakes}
\usetikzlibrary{decorations.markings}
\tikzstyle arrowstyle=[scale=1]
\tikzstyle directed=[postaction={decorate,decoration={markings,
    mark=at position .65 with {\arrow[arrowstyle]{stealth}}}}]
\tikzstyle endreversedirected=[postaction={decorate,decoration={markings,
    mark=at position 1.0 with {\arrow[arrowstyle]{stealth}}}}]
\tikzstyle enddirected=[postaction={decorate,decoration={markings,
    mark=at position 1.0 with {\arrow[arrowstyle]{stealth}}}}]

\tikzstyle reverse directed=[postaction={decorate,decoration={markings,
    mark=at position .65 with {\arrowreversed[arrowstyle]{stealth};}}}]

\newcommand{\Mathematica}[1]{}

\newcommand{\Eq}[1]{Eq.~(\ref{#1})}
\newcommand{\eq}[1]{(\ref{#1})}
\newcommand{\half}{\frac12}

\newcommand{\bea}{\begin{eqnarray}}
\newcommand{\eea}{\end{eqnarray}}
\newcommand{\beq}{\begin{equation}}
\newcommand{\eeq}{\end{equation}}
\newcommand{\be}{\begin{equation}}
\newcommand{\ee}{\end{equation}}

\newcommand{\rme}{\mathrm{e}}
\newcommand{\rmd}{\mathrm{d}}
\newcommand{\nn}{\nonumber}

\renewcommand{\epsilon}{\varepsilon}
\newcommand{\nott}[1]{}

\graphicspath{{figures/},{}}

\renewcommand{\paragraph}{\subsubsection*}

\begin{document}

\title{Depinning Transition of Charge-Density Waves: Mapping onto $O(n)$ Symmetric $\phi^4$ Theory
with $n\to -2$ and Loop-Erased Random Walks}
\author{Kay J\"org Wiese${}^{{1}}$ and Andrei A.\ Fedorenko${}^2$ }
\affiliation{$^1$\mbox{Laboratoire de Physique de l'Ecole normale sup\'erieure, ENS, Universit\'e PSL, CNRS, Sorbonne Universit\'e,} {Universit\'e Paris-Diderot, Sorbonne Paris Cit\'e, 24 rue Lhomond, 75005 Paris, France.}  \\
$^2$\mbox{Universit\'e de Lyon, ENS de Lyon, Universit\'e Claude Bernard, CNRS, Laboratoire
 de Physique, F-69342 Lyon, France} }


\begin{abstract}
Driven  periodic elastic systems such as charge-density waves (CDWs) pinned by impurities
show a nontrivial, glassy dynamical critical behavior. Their proper  theoretical description  requires the functional renormalization group. We show that
their critical behavior   close to the depinning transition is related to a much simpler
model, $O(n)$ symmetric $\phi^4$ theory in the unusual limit of $n\to -2$.
We demonstrate  that both theories yield identical results
to four-loop order and give both a perturbative and a nonperturbative proof of their equivalence.
As we show, both theories can be used to describe    {\em  loop-erased random walks} (LERWs), the trace of a random walk where loops are erased as soon as they are formed.
Remarkably, two famous models of non-self-intersecting
random walks, self-avoiding walks and  LERWs,  can both  be mapped onto
  $\phi^4$ theory, taken with formally $n=0$ and $n\to -2$ components.
This mapping allows us to compute the dynamic critical exponent of CDWs at the depinning transition and
the fractal dimension of LERWs in $d=3$ with unprecedented accuracy, $z(d=3)= 1.6243 \pm 0.001$,
in excellent agreement  with the  estimate  $z = 1.624\, 00 \pm 0.000\, 05$ of numerical simulations.
\end{abstract}

\maketitle

The model of periodic elastic manifolds driven by an external force through a disordered medium is
relevant for charge density waves (CDWs) in disordered solids
\cite{Gruner1988,FukuyamaLee1978,LeeRice1979}, flux-line lattices in the mixed state of disordered type-II superconductors (Bragg glass) \cite{BlatterFeigelmanGeshkenbeinLarkinVinokur1994,NattermannScheidl2000,LeDoussalGiamarchi1998,KleinJoumardBlanchardMarcusCubittGiamarchiLeDoussal2001},
and disordered Wigner crystals~\cite{Monceau2012,ReichhardtOlsonGronbech-JensenNori2001,ChitraGiamarchiLeDoussal1998}.
It has long been known   that even   weak disorder destroys the long-range translational  order and pins the elastic manifold
\cite{Larkin1970}.
Once the external driving
force $f$ exceeds a critical threshold force $f_{\rm c}$,
the manifold  undergoes a depinning transition to a sliding state. The   dynamics of the system in the vicinity of
this transition was studied both  numerically \cite{Middleton1992,MiddletonFisher1993,DuemmerKrauth2005,ScalaOliveLansacFilySoret2012,BustingorryKoltonGiamarchi2010},  and via field theory ~\cite{DSFisher1985,NarayanDSFisher1992b,NarayanDSFisher1992a,LeschhornNattermannStepanowTang1997,NattermannStepanowTangLeschhorn1992}. The latter
requires   the functional renormalization group (FRG). As
scaling arguments imply that the critical  behavior
of a disordered elastic manifold with short-range elasticity is dominated by disorder for
$d<d_{\mathrm{uc}}=4$,  any perturbative description breaks down on scales
larger than the {\em Larkin scale} \cite{LarkinOvchinnikov1979}. As a consequence,
one has to follow the renormalization of the whole disorder correlator which
{\blue develops a cusp}  at the Larkin scale.
{\red The appearance of this nonanalyticity in the running disorder correlator}
accounts for metastability and a finite  threshold force.
As the  corresponding FRG calculations are very involved,
  they have  only recently been extended to two-
\cite{LeDoussalWieseChauve2003,LeDoussalWieseChauve2002,ChauveLeDoussalWiese2000a}
and three-loop order \cite{WieseHusemannLeDoussal2018,HusemannWiese2017}.

\begin{figure}
\setlength{\unitlength}{1cm}\fboxsep0mm
{\begin{picture}(8.6,5.5)
\put(0,0){\includegraphics[width=\columnwidth]{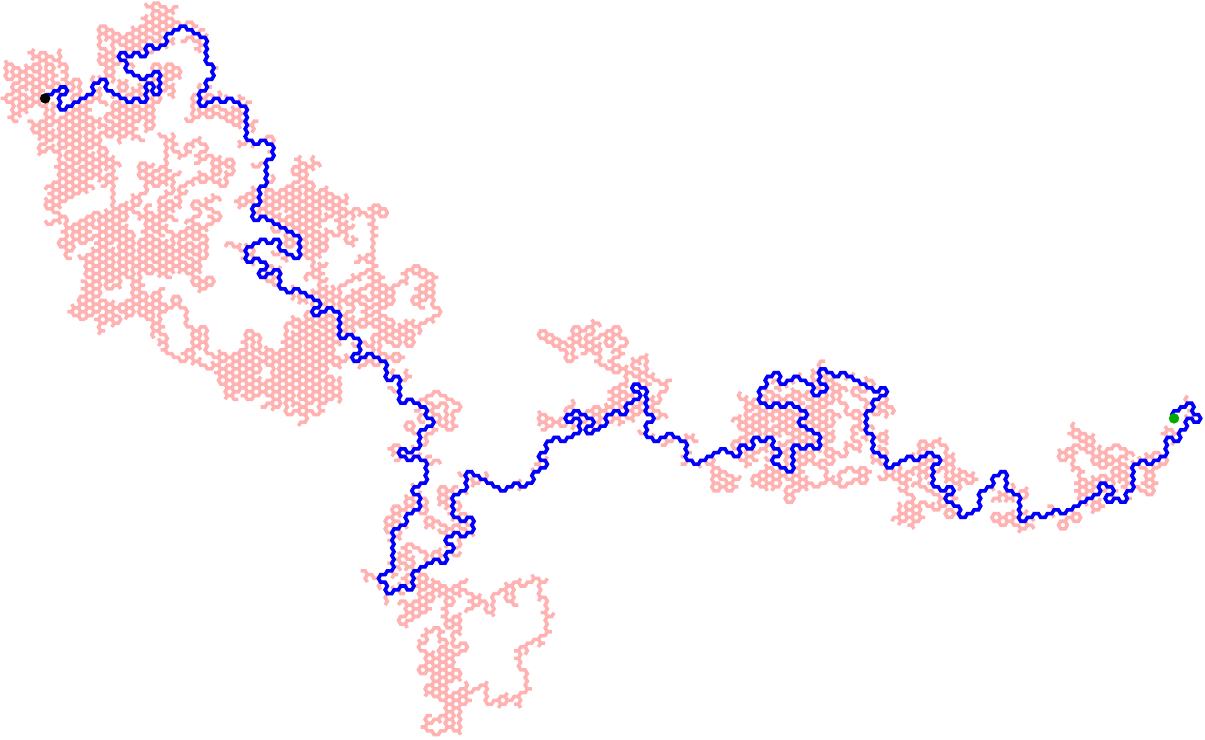}}
\put(3.5,3.1){\scalebox{0.7}{
{\begin{picture}(6.6,3.5)
\definecolor{kyellow}{rgb}{0.95,1,0.6}
\put(0,0){{\colorbox{kyellow}{\rule{72mm}{0mm}\rule{0mm}{35mm}}}}
\put(0.5,0.25){{\colorbox{red!30}{\rule{46mm}{0mm}\rule{0mm}{30mm}}}}
\put(1,1.9){{\colorbox{cyan!20}{\rule{36mm}{0mm}\rule{0mm}{10mm}}}}
\put(1.5,1.4){$\phi^{4}$-theory at $n\to -2$}
\put(1.3,0.8){1 boson and 2 fermions}
\put(2,0.5){(complex)}
\put(1.9,2.5){one family of}
\put(1.6,2.15){complex fermions}
\put(5.5,1.75){\begin{minipage}{1cm}
CDWs\\
at\\
depinning
\end{minipage}}
\end{picture}}}}
\end{picture}}
\caption{Trace of a LERW in blue, with the erased loops in red, on a 2D honeycomb lattice.
 (Inset) Nesting of the  different field theories for LERWs.}
\label{f:LERW}
\end{figure}
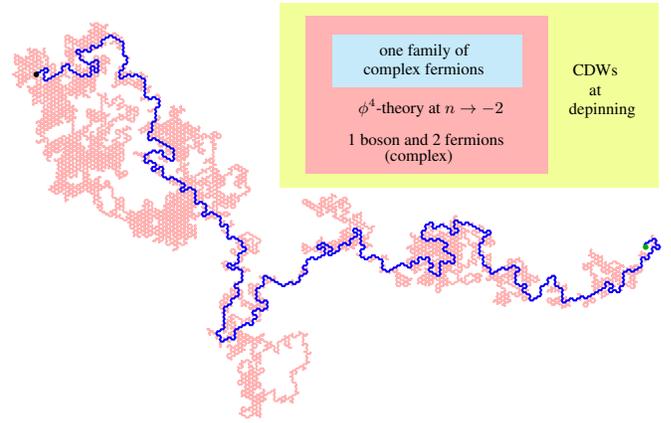

In the present Letter, we show that when the field  is periodic,   most properties are  described by a much
simpler field theory, namely, the $O(n)$ symmetric $\phi^4$ model with $n\to -2$.
This fact,  overlooked for   decades, drastically simplifies calculations
of the depinning transition, since $\phi^4$ theory
is  well known and its renormalization-group description does not require  the FRG.
We also prove that   both models  describe {\em loop-erased random walks}
(LERWs) in arbitrary dimension $d$.
In this Letter, we   outline the main ideas and results, while   details of the proof and
calculations are published elsewhere~\cite{WieseFedorenko2018}.

Random walks (RWs) {\em without} self-intersections play an important role in
mathematics, statistical physics and quantum field theory. The two widely encountered
models are {\em  self-avoiding walks} (SAWs) and  LERWs.
The SAW   describes long polymer chains  with self-repulsion
caused by excluded-volume effects. It can be defined as the uniform measure on all possible paths
of a given length without self-intersections. While the SAW  is   difficult
to analyze mathematically  rigorously, it was discovered by de Gennes \cite{DeGennes1972}
that its large-scale behavior can be extracted from the $O(n)$ symmetric
$\phi^4$ model in the unusual limit of $n \to 0$.
The LERW, which is intimately related to uniform spanning trees \cite{Majumdar1992,Dhar2006},
is a special case of the Laplacian RW \cite{LyklemaEvertszPietronero1986,Lawler2006}.
It is built from a  RW  by erasing any loop as soon as it is formed
\cite{Lawler1980}. A realization of a two-dimensional LERW  is shown in Fig.~\ref{f:LERW}.
Both models have a scaling limit in all dimensions, for instance,
the end-to-end distance $R$ scales with the RW length $\ell$
as $R \sim  \ell^{1/z} $, where $z$ is the fractal dimension~\cite{Kozma2007}.

Contrary to the SAW, the LERW has no obvious field-theory.
Three-dimensional LERWs have been studied only
numerically~\cite{GuttmannBursill1990,AgrawalDhar2001,Grassberger2009,Wilson2010}.
In two dimensions,  LERWs can be described by the radial
Schramm-Loewner evolution with parameter $\kappa=2$, also known as $\mathrm{SLE}_2$
\cite{Schramm2000,LawlerSchrammWerner2004}. It  predicts a fractal dimension  $z_{\rm LERW}(d=2)=5/4$,
which is clearly different from that of SAWs  $z_{\rm SAW}(d=2) = 4/3$.
 Coulomb-gas techniques link this to the
2D $O(n)$ model at $n\to -2$  which is a conformal field theory with  central charge
$c=-2$ \cite{Nienhuis1982,Duplantier1992}.  We show below
that the equivalence between LERWs and  $O(n)$ symmetric $\phi^4$ theory at $n\to -2$ holds in any dimension $d$.

In \cite{FedorenkoLeDoussalWiese2008a} it was   conjectured that the field theory of the depinning transition
of CDWs pinned by disorder is a field theory for LERWs.
This statement was based     on the conjecture of Narayan and Middleton \cite{NarayanMiddleton1994}
that  pinned CDWs  can be mapped onto the Abelian sandpile model. The connection of the latter with
uniform spanning trees, and thus with LERWs, is well established \cite{Dhar2006}.
The two-loop predictions  of \cite{FedorenkoLeDoussalWiese2008a} agree with   rigorous mathematical bounds,
and have been tested against numerical simulations at the upper critical dimension $d_{\mathrm{uc}}=4$~\cite{Grassberger2009}, where it was found that they correctly reproduce the leading  and subleading logarithmic corrections.

If this conjecture holds, then the  $\phi^4$ theory at $n\to -2$ has to reproduce the FRG picture for CDWs,
at least  for observables  related to LERWs.
Below we prove that the $\beta$ function  and the critical exponents $z$, $\nu=1/2$, and $\eta=0$
coincide for these theories. This is done by using a perturbative analysis of diagrams, non-perturbative supersymmetry techniques,
and an explicit four-loop calculation for both models.
However, this does not mean that the theories are identical, since one theory can have observables
absent in the other.  For instance,
at   depinning,  CDWs exhibit avalanches \cite{NarayanMiddleton1994,RossoLeDoussalWiese2009,KasparMungan2013}, which are seemingly absent in the  $\phi^4$ theory.
We claim that in the {\em sector} in
which we can compare the two theories, they agree (see inset of Fig.~\ref{f:LERW}).

Before we demonstrate the relation between   CDWs and the
$n$-component $\phi^{4}$ theory at $n\to -2$, we outline   how the latter
can be used to study LERWs in arbitrary dimension $d$.
First of all, it is
convenient to rewrite the $\phi^{4}$ theory in terms of $N=n/2$ complex bosons
$\Phi$, with action
\be\label{theory2}
{\cal S}[\vec \Phi] := \!\! \int_x \!  \nabla \vec \Phi^*(x)
{\nabla  \vec\Phi}(x)  +  {m^2}  \vec \Phi^{*}(x)  \vec \Phi(x) + \frac{g}{  2}  [\vec \Phi^{*}(x) \vec \Phi(x)]^2.
\ee
It is  known perturbatively  that for $N=-1$ the full two-point correlation function $\left< \Phi_i^*(x) \Phi_j(x')\right>$
reduces to the free-theory value independent of $g$ \cite{Zinn-JustinBook2,KleinertSchulte-FrohlindeBook,KleinertNeuSchulte-FrohlindeLarin1991,KompanietsPanzer2017}. It can be   proven  nonperturbatively by mapping onto complex fermions.
Indeed, in  Feynman diagrams for a bosonic $\Phi^4$ theory each loop carries a factor of $N$.
In a fermionic $\Phi^4$ theory with $M$ fermions,  a closed fermion loop carries a factor of $-M$,  so that a theory
with $N$ bosons is equivalent to a theory with $N+M$ bosons and $M$ fermions, where $N$ and $M$ can be continued to arbitrary real numbers. In particular, $N=-1$ corresponds to $M=1$,  where the term quartic in
fermionic fields vanishes, proving nonrenormalization of the propagator.

{\red
We now sketch the equivalence, referring to the Supplemental Material~\cite{Supplemental} for details and an alternative  proof based on Ref.~\cite{SapozhnikovShiraishi2018}.
In Fourier space, the two-point correlator $\left< \Phi_i^*(k) \Phi_j(-k)\right>$
can be viewed as the Laplace transform of the $k$-dependent Green's function for a RW.
It is convenient to draw the trajectory of the RW in blue and, when it hits itself, color
the emerging loop in red instead of erasing  it.
Going to the lattice and studying configurations with exactly one self-crossing, the contributions from perturbation theory are
}
{\setlength{\unitlength}{1cm}
\bea \label{eq-RW}
&&
{\parbox{2.25\unitlength}{\begin{picture}(2.15,1)\put(0.15,0){\includegraphics[width=2\unitlength]{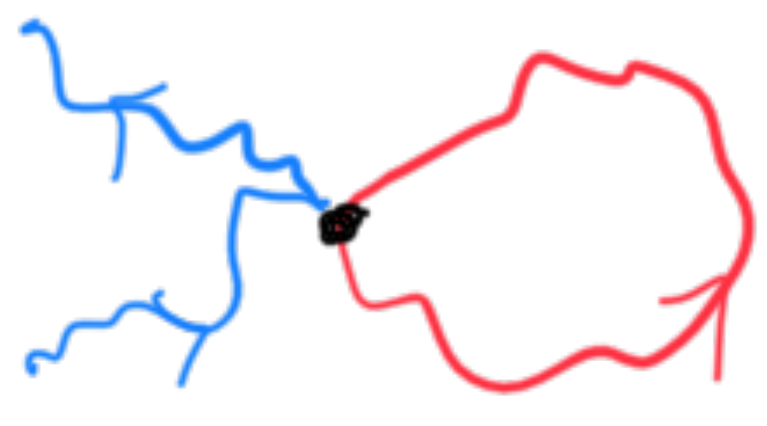}}
\put(0.,0){$\scriptstyle x$}
\put(1.0,0.7){$\scriptstyle y$}
\put(0.45,0){\scriptsize 1}
\put(1.9,0.5){\scriptsize 2}
\put(0.45,0.925){\scriptsize 3}
\put(0.03,0.95){$\scriptstyle x^{\prime}\ $}
\end{picture}}}\\
&&\longrightarrow
{\parbox{1.1cm}{{\begin{tikzpicture}
\coordinate (v1) at  (0,1.25) ; \coordinate (v2) at  (0,-.25) ;  \node (x) at  (0,0)    {$\!\!\!\parbox{0mm}{$\raisebox{-3mm}[0mm][0mm]{$\scriptstyle x$}$}$};
\coordinate (x1) at  (0.5,0);\coordinate (y) at  (1.5,0.5); \coordinate (y1) at  (0.5,1) ;\node (v) at  (0,1)    {$\!\!\!\parbox{0mm}{$\raisebox{1mm}[0mm][0mm]{$\scriptstyle x^{\prime}$}$}$};
\fill (x) circle (1.5pt);
\fill (v) circle (1.5pt);
\draw [blue] (x) -- (x1);
\draw [blue] (y1) -- (v);
\draw [blue,directed](0.5,0) arc (-90:90:0.5);
\end{tikzpicture}}}}~
-g
{\parbox{1.6cm}{{\begin{tikzpicture}
\node (v1) at  (0,1.25){} ;
\node (v2) at  (0,-.25){} ;
\node (x) at  (0,0)    {$\!\!\!\parbox{0mm}{$\raisebox{-3mm}[0mm][0mm]{$\scriptstyle x$}$}$};
\coordinate (x1) at  (1,0) ;\coordinate (y) at  (1.5,0.5); \coordinate (y1) at  (1,1);\node (v) at  (0,1)    {$\!\!\!\parbox{0mm}{$\raisebox{1mm}[0mm][0mm]{$\scriptstyle x^{\prime}$}$}$};
\fill (x) circle (1.5pt);
\fill (v) circle (1.5pt);
\fill (x1) circle (1.5pt);
\fill (y1) circle (1.5pt);
\node (y11) at  (1,0)    {$\!\!\!\parbox{0mm}{$\raisebox{-3mm}[0mm][0mm]{$\scriptstyle y$}$}$};
\node (y22) at  (1,1)    {$\!\!\!\parbox{0mm}{$\raisebox{1mm}[0mm][0mm]{$\scriptstyle y$}$}$};
\draw [blue,directed] (x) -- (x1);
\draw [blue,directed] (y1) -- (v);
\draw [blue,directed](1,0) arc (-90:90:0.5);
\draw [dashed] (x1) -- (y1);
\end{tikzpicture}}}}~
-g N
{\parbox{2.6cm}{{\begin{tikzpicture}
\node (v1) at  (0,1.25){} ;
\node (v2) at  (0,-.25){} ;
\node (x) at  (0,0)    {$\!\!\!\parbox{0mm}{$\raisebox{-3mm}[0mm][0mm]{$\scriptstyle x$}$}$};
\node (y1) at  (.8,0.5)    {$\!\!\!\parbox{0mm}{$\raisebox{-1mm}[0mm][0mm]{$\scriptstyle y$}$}$};
\node (y2) at  (1.7,.5)    {$\!\!\!\parbox{0mm}{$\raisebox{-1mm}[0mm][0mm]{$\scriptstyle y$}$}$};
\coordinate (x1) at  (0.5,0) ;\coordinate (y) at  (2.5,0.5);\coordinate (y1) at  (0.5,1);\coordinate (y2) at  (1.5,1) ; \node (v) at  (0,1)    {$\!\!\!\parbox{0mm}{$\raisebox{1mm}[0mm][0mm]{$\scriptstyle x^{\prime}$}$}$};
\coordinate (h1) at  (1,0.5) ;
\coordinate (h2) at  (1.5,0.5) ;
\fill (x) circle (1.5pt);
\fill (v) circle (1.5pt);
\fill (h1) circle (1.5pt);
\fill (h2) circle (1.5pt);
\draw [blue,directed] (x) -- (x1);
\draw [blue,directed] (y1) -- (v);
\draw [blue](0.5,0) arc (-90:90:0.5);
\draw [red,directed](1.5,0.5) arc (-180:180:0.5);
\draw [dashed] (h1) -- (h2);
\end{tikzpicture}}}}  \!\!\nn
\eea}{\red
The  first line is a graphical representation of the  RW  used to  construct a SAW or LERW.  It starts at $x$ and ends in $x^\prime$, passing
through the  segments numbered 1--3. By assumption,  it crosses   once at point $y$, but nowhere else.
The second line contains  all  one-loop diagrams of $\Phi^4$ theory.
De~Gennes  \cite{DeGennes1972} showed that setting $N\to 0$ yields the perturbative expansion of SAWs, a fact that can also be proven  algebraically~\cite{Zinn-JustinBook2}. In our formulation, the idea of the proof is as follows:
As we consider configurations with exactly one self-intersection, and since we are working on a lattice, the choice $g=1$  cancels the first two terms, while the last one is absent at $N=0$. Thus,  there is no  configuration with a self-intersection for SAWs. Now consider $g=1$ and $N \to -1$, for which the first two and last two terms cancel.
This implies that the free  propagator can be rewritten as the last diagram, which has the advantage to distinguish between red and blue parts of the trace, as long as the limit of $N\to -1$ is not yet taken.
The final step is to pass to the field theory. The latter has a $\beta$ function
with an attractive fixed point $g^{*}$ governing the large-distance behavior, implying that the   choice $g=1$ taken above can be relaxed to an arbitrary $g>0$.

What we need now is an operator that measures the length of the blue backbone in (\ref{eq-RW}). This is achieved by  the \textit{crossover operator}~\cite{Amit,Kirkham1981,ShimadaHikami2016},
\be\label{calO}
{\cal O}(y) :=   \Phi^{*} _{1}(y) \Phi _{1}(y)
- \Phi^{*} _{2}(y) \Phi  _{2}(y)  \ .
\ee
It checks whether point $y$ is part of the {\em blue} trace, as it vanishes in a {\em red} loop.
The fractal dimension $z$ of a LERW is extracted from the length of the  blue part via
\be\label{10}
 {\left< \int_{{y}} {\cal O}(y) \right>}
 \sim m^{-z}\ .
\ee}%
We now turn back to   CDWs which in the presence of disorder can be described by the Hamiltonian~\cite{NattermannScheidl2000,LeDoussalWieseChauve2003}
\begin{equation}\label{Hamiltonian}
  \mathcal{H} = \int_x \left\{ \frac{1}2 \Big[\nabla u(x)\Big]^2 +\frac {m^{2}}2 (u(x)-w)^{2} + V\big(x,u(x)\big)  \right\},
\end{equation}
where  $F(x,u)=-\partial_u V(x,u)$ is a random Gaussian force with zero mean and variance
$\overline{F(x,u)F(x',u')}=\Delta(u-u')\delta^d(x-x')$. The function $\Delta(u)$ is   even
with period $1$. The overdamped dynamics of CDWs  is given
by the equation of motion~$ \partial_t u(x, t) = -\delta \mathcal{H}[u]/ \delta u(x,t)$ \cite{NarayanDSFisher1992a}.
{\red Considering the system driven by increasing $w$ \cite{LeDoussalWiese2012a}, which means that the driving force $f$
fluctuates around its {\em self-organized} critical value $f_{\rm c}$
we arrive at the dynamic field theory \cite{LeDoussalWieseChauve2002,ChauveLeDoussalWiese2000a}}
\bea\label{CDW-action}
{\cal S}^{\rm CDW} &=& \int_{x,t} \tilde u (x,t)  (\partial_{t}-\nabla^{2}+m^2)  [ u
(x,t)-w]  \\
&& -\half \int_{x,t,t'} \tilde u (x,t)\tilde u
(x,t')  \Delta \big(u (x,t)-u (x,t')\big).~~~~~~\nn
\eea
The statistical tilt symmetry implies nonrenormalziation of the gradient and mass terms, equivalent to exponents $\nu=1/2$ and $\eta=0$ in the $\phi^4$ model at $n\to -2$.

One checks that in the theory (\ref{CDW-action}) all Taylor  coefficients in the
expansion of $\Delta(u)$ at $u=0$ are relevant coupling constants  for $d<4$ so that one has to follow renormalization of the whole function. This can be achieved by using the FRG \cite{DSFisher1985,NarayanDSFisher1992a,NarayanDSFisher1992b,LeschhornNattermannStepanowTang1997,NattermannStepanowTangLeschhorn1992,LeDoussalWieseChauve2003,LeDoussalWieseChauve2002,ChauveLeDoussalWiese2000a}.
The flow equation to one-loop order is
\begin{equation}
-m \partial_{m} \Delta (u) = \varepsilon \Delta (u)
- \frac12 \frac{\rmd^2}{\rmd u^2} [\Delta(u)-\Delta(0)]^2, \label{eq:FRG}
\end{equation}
where $\varepsilon =4-d$.
The analysis of the FRG flow shows that the fixed point (FP) with period $1$ has
the form $\Delta(u) = \Delta(0) - \frac g{ 2} u(1-u)$ for $u \in [0,1]$
with a cusp at the origin.
In the absence of higher-order terms in $u$, the renormalization group flow closes in the   space of polynomials of degree $2$, and for the quadratic term one is left with the renormalization of a single coupling constant $g$. This form of the FP has been confirmed explicitly to three-loop order
and presumably holds to all orders \cite{WieseHusemannLeDoussal2018,HusemannWiese2017}.

In order to   connect  to the $\Phi^4$ theory introduced above, let us use supersymmetry to  average over disorder \cite{ParisiSourlas1979,ParisiSourlas1982,Wiese2004}. The validity of  this method at depinning is justified by the fact that
the periodic FPs describing  depinning and equilibrium have the same value of
$g$ and differ only by $\Delta(0)$. At equilibrium the FP is potential, i.e.,
$\int_{-\infty}^{\infty}\rmd u\,\Delta(u)=0$, and thus $g$ also determines $\Delta(0)$. At depinning   $g$
is not enough to get the whole two-point function, and   some information  is absent.
The disorder average of any observable ${\cal O}[u_i]$ is \cite{ParisiSourlas1979,ParisiSourlas1982,Wiese2004}
\begin{align}\label{su5}
&\overline {{\cal O}[u_i]}  = \int\prod_{a=1}^2 {\cal D}[{\tilde u_{a} }] {\cal D}[{
u_{a} }] {\cal D}[{\bar \psi_{a} }] {\cal D}[{\psi}_{a}]\, {{\cal O}[u_i]}  \\
& ~~~\times
\!\overline{\,\exp\!\left[-{ \int_{x}\tilde u_{a} (x)\frac{\delta{\cal H}[{u_{a} }]} {\delta {u_{a} } (x) }+\bar \psi_{a}(x)\frac{\delta^{2}
{\cal H}[{u_{a} } ] }{\delta {u_{a} } (x)\delta {u_{a} }
(y) } \psi_{a} (y) } \right]}. \nn
\end{align}
Here the integral over the auxiliary bosonic fields $\tilde u_a$ implies that $u_a$ is at a  minimum of $\cal H$,  while
the integrals over fermionic fields $\bar \psi_a$ and $\psi_a$
cancel the functional determinant appearing in the integration over $u_{a}$.

{\red It is known that direct application of this method with one copy
fails beyond the Larkin length,  leading to the so-called dimensional reduction \cite{ParisiSourlas1979,ParisiSourlas1982}.
The key point is that we introduced two copies $a=1,2$ of the system in (\ref{su5})
to get access to the second cumulant of the disorder distribution that
we want to renormalize. As was shown in Ref.~\cite{Wiese2004}, one   recovers the FRG flow equation \eq{eq:FRG} of the statics, which in turn  leads to the appearance of a cusp in the running disorder correlator at the Larkin scale, thus avoiding dimensional reduction. It can also be viewed as a breaking of supersymmetry.}

Introducing center-of-mass coordinates
\begin{equation}
u_{1,2}(x) = u(x) \pm  \frac12 \phi(x), \ \ \ \
\tilde u_{1,2}(x) = \frac12 \tilde u(x) \pm  \tilde \phi(x)\ ,
\end{equation}
 the effective action becomes after some cumbersome but straightforward calculation  shown in the Supplemental Material~\cite{Supplemental}
\begin{eqnarray}
\label{CDW=phi4}
{\cal S} &=&   \int_{x} \tilde \phi(x) (-\nabla^2 +m^2)\phi(x)+ \tilde  u(x) (-\nabla^2 +m^2) u(x)
\nn\\
&+&   \sum_{a=1}^2\bar \psi_{a} (x)
(-\nabla^{2}+m^2)\psi_{a} (x) \nonumber \\
&+&
 \frac g 2\tilde u(x) \phi(x)\!\left[\bar \psi_2(x)\psi_2(x) -\bar \psi_1(x)\psi_1(x) -\frac14 \tilde u(x)   \phi(x)\right] \nonumber \\
&+& \frac g 2\left[ \tilde \phi(x)\phi(x)+\bar \psi_1(x)\psi_1(x) +\bar \psi_2(x)\psi_2(x) \right]^2 \ .
 \end{eqnarray}
It is easy to check, that while  $u(x)$ and $\tilde{u}(x)$    have nontrivial expectations,
the terms depending on them (the second term in the first line, and the third line) do
not contribute to the  renormalization of $g$ and thus can be dropped. What   remains in action \eq{CDW=phi4} is   a $\Phi^4$-type theory with one ($N=1$) complex boson and two ($M=2$) complex fermions.
As we showed above, this can equivalently be viewed as complex $\Phi^4$ theory with $N\to -1$, or real $\phi^4$ theory with $n\to -2$.
{\red  We thus proved that   both models have the same effective coupling $g$, and thus the same   $\beta$ function for $g$.
This allows us to reconstruct $\Delta(u)$ in the statics and up to the constant $\Delta(0)$ also at depinning.
 }

{
We show now that this relation between the two models allows one to determine the dynamic exponent $z$ at depinning.
The dynamic theory has an additional  renormalization of friction or time, which shows up in corrections to the term $\int_{x,t} \tilde u(x,t) \dot u(x,t)$ in action (\ref{CDW-action}).
Using this action  to construct all diagrams  in which one field $\tilde u$ and one field $u$ remain,
the latter has the form $u(x,t)-u(x,t')$ and can be expanded as $\dot u(x,t) (t-t')$.
The time difference, when appearing in the expression for a diagram together with a response
function given in Fourier by $ R(k,t) = \Theta(t) \rme^{-t\, (k^2+m^2)}$,  can be treated as an insertion
of an additional point  into the line for the latter using the relation
\be
 t R(k,t) =  \int_0^t \rmd{t'} R(k,t') R(k,t-t') \label{eq-R-relation} \ .
\ee
One can check perturbatively  that the diagrams renormalizing the term $\tilde u (x,t)  \partial_{t}
u (x,t) $  in the CDW action~(\ref{CDW-action}) reduce
to the two-point function of  model (\ref{theory2}) with an insertion
of the crossover operator~(\ref{calO}).
This   identifies the dynamic exponent of CDWs at depinning with the
crossover exponent of the $\Phi^4$ theory. Let us demonstrate this on the example of
the one-loop dynamic  diagram
\begin{eqnarray} \label{eq:depinning-insertion}
{\parbox{1.6cm}{{\begin{tikzpicture}
\node (x) at  (0,0)  {$\!\!\!\parbox{0mm}{$\raisebox{-3mm}[0mm][0mm]{$\  \scriptstyle t$}$}$};
\node (y) at  (1.5,0)  {$\!\!\!\parbox{0mm}{$\raisebox{-3mm}[0mm][0mm]{$\ \ \scriptstyle t'$}$}$};
\fill (x) circle (1.5pt);
\fill (y) circle (1.5pt);
\draw [dashed] (x) -- (y);
\draw [blue,directed](y) arc (0:180:0.75);
\coordinate (x1) at  (0,0.75);
\coordinate (y1) at  (1.5,-0.4);
\coordinate (y11) at  (1.5,0.0);
\draw [blue,directed] (0,0) -- (x1);
\draw [blue,directed] (y1) -- (y11);
\end{tikzpicture}}}}
\hspace{3mm} \longrightarrow \hspace{3mm}
{\parbox{1.6cm}{{\begin{tikzpicture}
\coordinate (x) at  (0,0);
\coordinate (y) at  (1.5,0) ;
\coordinate (z) at  (0.75,0.75);
\draw[red,snake=snake,segment length=3pt] (z)--(0.75,1.1) ;
\fill (x) circle (1.5pt);
\fill (y) circle (1.5pt);
\fill (z) circle (1.5pt);
\draw [dashed] (x) -- (y);
\draw [blue,directed](y) arc (0:90:0.75);
\draw [blue,directed](z) arc (90:180:0.75);
\end{tikzpicture}}}} \ .
\end{eqnarray}
The wavy line is   the crossover operator defined in \Eq{calO}.
Using a short-time expansion,  the lhs \ of Eq.~(\ref{eq:depinning-insertion}) is evaluated to~\cite{LeDoussalWieseChauve2002}
\begin{equation}
   \int_{x, t,t'} \tilde{u}(x,t) \left [ \Delta'(0^+) + \Delta''(0) (t-t') \dot{u}(x,t) \right ] R_{0,t-t'},\end{equation}
where $R_{x,t}$  is the response function in real space.
The first term  $\sim \Delta'(0^+)$ renormalizes  the critical force, while the second one corrects the friction. Using  relation (\ref{eq-R-relation})
and integrating over times,   the resulting expression  is the one-loop diagram of   $\Phi^4$ theory for the observable  \eq{calO}, i.e.,\ the rhs \ of \Eq{eq:depinning-insertion}.
Following this strategy, we checked that this property persists up to four-loop order.
This can be proven graphically to all orders \cite{WieseFedorenko2018}.
}

We  generated all diagrams contributing to {${\cal O}(y)$} at five-loop order,   and  to the  renormalization of the coupling constant at four-loop order, using the diagrams computed in a massless minimal subtraction scheme in \cite{KleinertNeuSchulte-FrohlindeLarin1991,KompanietsPanzer2017}.
This yields for the dynamical exponent $z$ of CDWs at depinning in dimension $d=4-\epsilon$, equivalent to the   fractal dimension $d_{\rm f}$ of LERWs in the same dimension,
\begin{eqnarray} \nn
\!\!\!z &=&  2-\frac{\varepsilon }{3}- \frac{\varepsilon^2}{9}
+ \bigg[\frac{2 \zeta (3)}{9}-\frac{1}{18}\, \bigg]\, \varepsilon^3  \\
\nonumber
&& - \bigg[\,\frac{70 \zeta (5)}{81} -\frac{\zeta(4)}{6} -\frac{17 \zeta (3) }{162}
   +\frac{7}{324}\, \bigg]\, \varepsilon^4 \nn\\
   &&     + \bigg[
   \frac{121 \zeta (3)}{972} -\frac{8 \zeta (3)^2}{81}+\frac{17
   \zeta (4)}{216}-\frac{103 \zeta (5)}{243}-\frac{175 \zeta
   (6)}{162}
  \nn\\
   &&   ~~~~+\frac{833 \zeta (7)}{216}-\frac{17}{1944} \, \bigg]\, \varepsilon^5 + \, {\cal O}(\varepsilon^6), \label{eq:On-24} \ \ \ \ \
\end{eqnarray}
where  $\zeta (s)$ is the Riemann zeta function.
This result agrees with the dynamic critical exponent of CDWs at depinning
computed using FRG to two- \cite{LeDoussalWieseChauve2002} and four-loop order \cite{WieseFedorenko2018}; the four-loop result for the crossover exponent of the $O(n)$ symmetric  $\phi^4$ theory computed  in Ref.~\cite{Kirkham1981}, setting $n\to -2$, and its extension to six-loop order \cite{KompanietsWiese2019}.
Using Borel resummation of the latter yields   $z = 1.244\pm 0.01$
in $d=2$, where the exact value is $z=5/4$  \cite{Schramm2000,LawlerSchrammWerner2004}, and
$z(d=3)= 1.6243 \pm 0.001$.
This can be compared to the most precise numerical simulations to date by Wilson \cite{Wilson2010},
$z(d=3) = 1.624\ 00 \pm 0.000\ 05$.

To summarize, we showed that CDWs at   depinning  are equivalent to the $O(n)$-symmetric
$\phi^{4}$ theory with  $n\to -2$, and that  both  field theories describe LERWs.
We gave both a perturbative proof of this equivalence and a proof based on  supersymmetry. This was    checked    by an explicit four-loop calculation.
Using the $O(n)$ symmetric $\phi^4$ theory  we   calculated the dynamic critical exponent for CDWs at depinning and the fractal dimension of LERWs to fifth order in $\varepsilon=4-d$, in excellent agreement with known numerical results.
Our findings are   surprising, since a  simple  $\phi^4$ theory  allows one to obtain the FRG fixed point of CDWs, which is a glassy disordered system.
However, it does not provide
all information about pinned CDWs, for instance, the two-point dynamic correlation function.
Our understanding is that   both field theories are not isomorphic,
but when restricted to the same {\em physical sector} make the same predictions.
This opens a path to eventually tackle other systems, which currently necessitate the FRG, such as  random-field magnets  \cite{Feldman2002,TarjusTissier2004,LeDoussalWiese2005b}, using a simpler effective field theory.

Our results  provide a strong support for the Narayan-Middleton conjecture~\cite{NarayanMiddleton1994} that CDWs pinned by disorder can be mapped onto the Abelian sandpile model and on LERWs~\cite{FedorenkoLeDoussalWiese2008a}. As a consequence,  the dynamic critical exponent
of a 2D CDW at depinning is exactly $z(d=2)=5/4$.
Remarkably, while CDWs at depinning map onto Abelian sandpiles, disordered elastic interfaces at depinning   map onto Manna sandpiles \cite{LeDoussalWiese2014a,Wiese2015}. Thus,  each main universality class at depinning corresponds to a specific sandpile model.

Finally, the  mapping of $\phi^4$-theory at $n\to -2$ onto LERWs provides not only the fractal dimension
of the latter, but also the  correction-to-scaling exponent $\omega$. We propose to measure it in simulations by erasing loops   with  probability $p<1$. Its $\varepsilon$-expansion at six-loop order~\cite{KompanietsPanzer2017} is only slowly converging, and  we estimate $\omega = 0.83 \pm 0.01$.

\textit{Acknowledgements.}
It is a pleasure to thank E.~Br\'ezin, J.~Cardy, F.~David,  K.~Gawedzki, P.~Grassberger, J.~Jacobsen,
  M.V.~Kompaniets, A.~Nahum, S.~Rychkov, D.~Wilson, and J.~Zinn-Justin for valuable discussions.
  A.A.F.\ acknowledges   support from the ANR Grant  No. ANR-18-CE40-0033 (DIMERS).

\renewcommand{\doi}[2]{\href{http://dx.doi.org/#1}{#2}}
\newcommand{\arxiv}[1]{\href{http://arxiv.org/abs/#1}{#1}}


\begin{widetext}
\mbox{}\\[6mm]

\newpage
\centerline {\Large \bf
  SUPPLEMENTAL MATERIAL}

\appendix
\hypertarget{sec:A}{}

\section{A. Details on the mapping of LERW on $O(n=-2)$ model} \label{sec:LERW1}

We use that in Fourier space the 2-point correlator of the $O(N)$ $\Phi^4$ model
can be viewed as the Laplace transform of the $k$-dependent Green function for a RW.
\be\label{phi-prop}
\left< \Phi_i^*(k) \Phi_j(-k)\right> ={\blue \delta_{{ij}}} \int_{0}^{\infty}\rmd t\, \rme^{-m^{2}t } \times \rme^{-k^{2 } t }\ .
\ee
Here $t\ge \ell$ is the time of the RW used to construct a  LERW of length $\ell$,
which scales as $\ell\sim t^{z/2}\sim m^{-z}$, and $z$ is the fractal dimension of the LERW.
It is convenient to draw the trajectory of the RW in blue, and when it hits itself color
the emerging loop in red instead of erasing  it.
We claim that we can deduce the statistics of these {\em colored} RWs  from the $\phi^{4}$ theory.

To render this construction more transparent, we make the argument for self-avoiding polymers ($N=0$), and loop-erased random walks ($N\to -1$) at the same time. The former equivalence is  known since de~Gennes  \cite{DeGennes1972}, and can be proven algebraically (see e.g.~\cite{Zinn-JustinBook2}), the latter is what we wish to establish here.
To be specific about UV cutoffs, we put the system on a lattice. The indicator function of a self-intersection is then 1 if the paths have a common vertex, and zero otherwise.

Consider a specific RW with  $s=1$ self-intersections,
{\setlength{\unitlength}{1cm}
\bea
&&
{\parbox{2.25\unitlength}{\begin{picture}(2.15,1)\put(0.15,0){\includegraphics[width=2\unitlength]{trace1}}
\put(0.,0){$\scriptstyle x$}
\put(1.0,0.7){$\scriptstyle y$}
\put(0.45,0){\scriptsize 1}
\put(1.9,0.5){\scriptsize 2}
\put(0.45,0.925){\scriptsize 3}
\put(0.03,0.95){$\scriptstyle x'$}
\end{picture}}} \ \ 
\longrightarrow \ \
{\parbox{1.1cm}{{\begin{tikzpicture}
\coordinate (v2) at  (0,-.25); \coordinate (v1) at  (0,1.25);   \node (x) at  (0,0)    {$\!\!\!\parbox{0mm}{$\raisebox{-3mm}[0mm][0mm]{$\scriptstyle x$}$}$};
\coordinate (y) at  (1.5,0.5); \coordinate (x1) at  (0.5,0); \coordinate (y1) at  (0.5,1) ;\node (v) at  (0,1)    {$\!\!\!\parbox{0mm}{$\raisebox{1mm}[0mm][0mm]{$\scriptstyle x'$}$}$};
\fill (v) circle (1.5pt);
\fill (x) circle (1.5pt);
\draw [blue] (y1) -- (v);
\draw [blue] (x) -- (x1);
\draw [blue,directed](0.5,0) arc (-90:90:0.5);
\end{tikzpicture}}}}~ \ \
-g \ \
{\parbox{1.6cm}{{\begin{tikzpicture}
\node (v2) at  (0,-.25){} ;
\node (v1) at  (0,1.25){} ;
\node (x) at  (0,0)    {$\!\!\!\parbox{0mm}{$\raisebox{-3mm}[0mm][0mm]{$\scriptstyle x$}$}$};
\coordinate (x1) at  (1,0) ;\coordinate (y) at  (1.5,0.5); \coordinate (y1) at  (1,1);\node (v) at  (0,1)    {$\!\!\!\parbox{0mm}{$\raisebox{1mm}[0mm][0mm]{$\scriptstyle x'$}$}$};
\fill (x1) circle (1.5pt);
\fill (x) circle (1.5pt);
\fill (y1) circle (1.5pt);
\fill (v) circle (1.5pt);
\node (y11) at  (1,0)    {$\!\!\!\parbox{0mm}{$\raisebox{-3mm}[0mm][0mm]{$\scriptstyle y'$}$}$};
\node (y22) at  (1,1)    {$\!\!\!\parbox{0mm}{$\raisebox{1mm}[0mm][0mm]{$\scriptstyle y'$}$}$};
\draw [blue,directed] (x) -- (x1);
\draw [blue,directed] (y1) -- (v);
\draw [blue,directed](1,0) arc (-90:90:0.5);
\draw [dashed] (x1) -- (y1);
\end{tikzpicture}}}}~ \ \
-g N \ \
{\parbox{2.6cm}{{\begin{tikzpicture}
\node (v1) at  (0,1.25){} ;
\node (v2) at  (0,-.25){} ;
\node (x) at  (0,0)    {$\!\!\!\parbox{0mm}{$\raisebox{-3mm}[0mm][0mm]{$\scriptstyle x$}$}$};
\node (y1) at  (.8,0.5)    {$\!\!\!\parbox{0mm}{$\raisebox{-1mm}[0mm][0mm]{$\scriptstyle y'$}$}$};
\node (y2) at  (1.7,.5)    {$\!\!\!\parbox{0mm}{$\raisebox{-1mm}[0mm][0mm]{$\scriptstyle y'$}$}$};
\coordinate (x1) at  (0.5,0) ;\coordinate (y) at  (2.5,0.5);\coordinate (y1) at  (0.5,1);\coordinate (y2) at  (1.5,1) ; \node (v) at  (0,1)    {$\!\!\!\parbox{0mm}{$\raisebox{1mm}[0mm][0mm]{$\scriptstyle x'$}$}$};
\coordinate (h1) at  (1,0.5) ;
\coordinate (h2) at  (1.5,0.5) ;
\fill (x) circle (1.5pt);
\fill (v) circle (1.5pt);
\fill (h2) circle (1.5pt);
\fill (h1) circle (1.5pt);
\draw [blue,directed] (x) -- (x1);
\draw [blue](0.5,0) arc (-90:90:0.5);
\draw [blue,directed] (y1) -- (v);
\draw [dashed] (h1) -- (h2);
\draw [red,directed](1.5,0.5) arc (-180:180:0.5);
\end{tikzpicture}}}}\!\!\nn
\eea}
The first line is a graphical representation of the  RW used to  construct a SAW or LERW.  It starts at $x$ and ends in $x'$, passing
through the  segments numbered 1 to 3. By assumption  it crosses   once at point $y$, but nowhere else.
The second line contains  all  diagrams of $\Phi^4$ theory up to order $g^{s}$:
the first term is the free-theory result, proportional to $g^{0}$;   the second and third term $\sim g$ are the 1-loop perturbative corrections. The lattice point $y'$ of self-intersection in the $\phi^{4}$-interaction is summed over; this sum has exactly one non-vanishing term, namely when $y=y'$. The choice $g=1$   leads to a cancelation of the first two diagrams.

Let us first consider SAWs, i.e.\ $N=0$. Then the last term vanishes, and the equation   says that (for $g=1$) configurations with one self-intersection are absent from the partition function of SAWs.
Next consider $N=-1$: Then there are two cancelations, ({\em i}) between the first two terms, and ({\em ii}) between the last  two terms.
This shows two things: due to the second cancelation,   the propagator is that of the free theory.
Due to the first cancelation,  one can rewrite the drawing we started with as the last diagram.
The advantage of this  rewriting is that one  distinguishes between the backbone in blue, and the loop in red, as long as one keeps $N$ as a parameter.

Two remarks are in order:
First, our construction was done on the lattice, with {\em bare coupling} $g=1$.
As the renormalization group  tells us,  the effective coupling and the universal properties of the system are independent  of this choice.
Second, one has to prove this cancelation recursively for more than one self-intersection \cite{WieseFedorenko2018}.

We sketch a more mathematical proof here (details are given elsewhere):  \cite{SapozhnikovShiraishi2018} theorem 1.1 states  that
the union of a LERW and the loops from the {\em loop soup}
ensemble of oriented loops    intersecting  this LERW has the same law as a random walk. (All ensembles are conditioned to the inside of the unit ball).
Denote the loop-soup ensemble by LS(2).
Symbolically, we note
$
\mbox{RW} = \mbox{LERW} \oplus \mbox{LS}(2),
$ which we can rewrite as
\be
 \mbox{LERW} = \mbox{RW}  \ominus \mbox{LS}(2) = \mbox{RW}   \oplus \mbox{LS}(-2)\ .
\ee
Now use   that $ \mbox{LS}(-2)$ in our language is a free theory with $n= -2$, and that its partition function is identical to that of \Eq{theory2} at any $g$. Remains to identify the weight of intersections; this is the same choice of $g=1$ used above.

\section{B. Proof for the equivalance of $\phi^4$-theory at $N=-1$ and CDWs} \label{sec:LERW1}

A method to average over disorder using both bosonic and fermionic degrees of freedom was introduced in Ref.~\cite{ParisiSourlas1979,ParisiSourlas1982}. It is better known as
the supersymmetry method, even though supersymmetry may be broken, and is, as we will see below, indeed broken   beyond the Larkin scale.
The method allows one to write the disorder average of any observable ${\cal O}[u_i]$ as
\begin{align}\label{su5-bis}
&\overline {{\cal O}[u_i]}  = \int\prod_{a=1}^r {\cal D}[{\tilde u_{a} }] {\cal D}[{
u_{a} }] {\cal D}[{\bar \psi_{a} }] {\cal D}[{\psi}_{a}]\, {{\cal O}[u_i]}
 \overline{\,\exp\!\left[-{ \int_{x}\tilde u_{a} (x)\frac{\delta{\cal H}[{u_{a} }]} {\delta {u_{a} } (x) }+\bar \psi_{a}(x)\frac{\delta^{2}
{\cal H}[{u_{a} } ] }{\delta {u_{a} } (x)\delta {u_{a} }
(y) } \psi_{a} (y) } \right]} . \end{align}
The integral over $\tilde u_a$ ensures that  $u_a$ is at a minimum. $\bar \psi_a$ and $\psi_a$ are Grassmann variables, which compensate for the functional determinant appearing in the integration over $u$. As a result, the  partition function ${\cal Z}=1$.
The  effective action after averaging over disorder is  \cite{Wiese2004}
\begin{eqnarray}\label{su6a-bis}
\label{su6b-bis}
{\cal S}[\tilde u_a, u_{a},\bar \psi_{a}, \psi_{a}]
&=& \!
\sum_{a}\! \int_{x} \!\tilde u_{a} (x) (-\nabla^{2}{+}m^2) u_{a} (x)   + \bar \psi_{a} (x)
(-\nabla^{2}{+}m^2)\psi_{a} (x) \nonumber
\\
&-& \sum_{a,b} \int_x\Big[ \half \tilde u_{a} (x)\Delta \big(u_{a} (x)-u_{b}
(x)\big)\tilde u_{b} (x)
- \tilde u_{a} (x)
\Delta' \big(u_{a} (x)-u_{b} (x)\big) \bar \psi_{b} (x)\psi_{b} (x)\nonumber \\
&& \qquad~~~   -\half \bar \psi_{a} (x)\psi _{a} (x)\Delta''
\big(u_{a} (x)-u_{b} (x)\big)\bar \psi_{b} (x)\psi_{b} (x)  \Big]. \end{eqnarray}
This expression contains a sum over an arbitrary number  $r$ of replicas (copies).
To extract the correlations of the disorder, or formally its  second cumulant, one needs at least $r=2$ replicas.
If one were to use $r=3$ copies, one would in addition have access to the third cumulant of the disorder. Since we only need the second cumulant, and since this formulation is simpler, we now choose $r=2$.
Note that   the seminal work  \cite{ParisiSourlas1979}   focused   on $r=1$, which prevents one from extracting the second cumulant of the disorder.

Let us define  center-of-mass coordinates,
\begin{equation}
u_1(x) = u(x) + \frac12 \phi(x)\ ,  \qquad  u_2(x)= u(x) - \frac12  \phi(x)\ , \qquad
\tilde u_1(x) = \frac12 \tilde u(x) + \tilde \phi(x)\ ,  \qquad \tilde u_2(x)= \frac12  \tilde u(x) -  \tilde \phi (x)\ .
\end{equation}
This allows us to rewrite the action \eq{su6b-bis}   as
\begin{align}
\label{CDW=phi4bis-bis} {\cal S} =   \int_{x} &\tilde \phi(x) (-\nabla^2 +m^2)\phi(x)+ \tilde  u(x) (-\nabla^2 +m^2) u(x)
+   \sum_{a=1}^2\bar \psi_{a} (x)
(-\nabla^{2}+m^2)\psi_{a} (x)\nn\\
&+  \tilde \phi(x)^2 \Big[ \Delta\big(\phi(x)\big)-\Delta(0)\Big] -\frac14 \tilde u(x)^2\Big[ \Delta\big(\phi(x)\big)+\Delta(0)\Big]
+ \frac12 \tilde u(x) \Delta'\big(\phi(x)\big)\Big[\bar \psi_{2}(x) \psi_{2}(x)-\bar \psi_{1}(x) \psi_{1}(x)\Big] \nn\\& +  \tilde \phi(x) \Delta'\big(\phi(x)\big)\Big[\bar \psi_{2}(x) \psi_{2}(x)+\bar \psi_{1}(x) \psi_{1}(x)\Big]
+ \bar \psi_{2}(x) \psi_{2}(x) \bar \psi_{1}(x) \psi_{1}(x) \Delta''\big(\phi(x)\big)\ .
 \end{align}
Replacing
\be\label{AA}
\Delta(u)= \Delta(0)-\frac g2 u(1-u) \to  \Delta(0) + \frac g{ 2} u^{2}\ ,
\ee the action  takes the form
\begin{align}
\label{CDW=phi4-cis} {\cal S} =   \int_{x} &\tilde \phi(x) (-\nabla^2 +m^2)\phi(x)+ \tilde  u(x) (-\nabla^2 +m^2) u(x) -\frac{\Delta(0)}2 \tilde u(x)^{2}
+   \sum_{a=1}^2\bar \psi_{a} (x)
(-\nabla^{2}+m^2)\psi_{a} (x) \nonumber \\
&+
 \frac g 2\tilde u(x) \phi(x)\!\Big[\bar \psi_2(x)\psi_2(x) -\bar \psi_1(x)\psi_1(x) \Big]   - \frac g 8 \tilde u(x) ^2 \phi(x)^2
 \nonumber \\ &
+ \frac g 2\left[ \tilde \phi(x)\phi(x)+\bar \psi_1(x)\psi_1(x) +\bar \psi_2(x)\psi_2(x) \right]^2 \ .
 \end{align}
Note that the center-of-mass  position $u(x)$ does not appear in the interaction, only the field $\tilde u(x)$.   As a consequence, $u(x)$    does not participate in the renormalization of $g$, and the latter can be obtained by dropping the second and third  line of \Eq{CDW=phi4}. What remains  is a $\phi^4$-type theory with one complex boson $\phi$, and two complex fermions $\psi_{1}$ and $\psi_{2}$. It can equivalently be viewed as complex $\phi^4$-theory at $N\to -1$, or real $\phi^4$-theory at $n\to -2$. This proves the statements   made in the main text.

Note that if one were to include the term of order $g u$ in $\Delta(u)$, a term of the form $\tilde u(x) \sum_{{a=1}}^{2} \bar \psi_{a}(x) \psi_{a}(x)$ would appear, renormalizing $\Delta(0)$, and leading to a breaking of supersymmetry.

As we explained in the manuscript, the case $N=-1$ for the bosonic field  $\Phi$  corresponds to one flavour ($M=1$) of fermions with contact interactions given by the quartic term $g$. Since the quartic term vanishes for $M=1$ due to properties of Grassmann variables, one arrives at free fermions. Remarkably, one cannot extract the properties of CDWs or LERWs directly from free fermions, despite the fact that their partition function is related to the number of uniform spanning trees \cite{Wu1977}. As we showed, however, this can be done by studying  interacting fermions with $M$ flavors and taking the limit of $M \to 1$   at the end. This   trick renders the system quasi-interacting rather than free, with a  non-trivial renormalization of $g$ which encodes the properties of CDWs and LERW, even though the two-point function is not corrected.

\end{widetext}

\end{document}